\documentclass[prl, twocolumn, preprintnumbers,amsmath,amssymb,superscriptaddress,longbibliography]{revtex4-2}
\usepackage{graphicx}
\usepackage{dcolumn}
\usepackage{bm}
\usepackage{subfigure}
\usepackage{multirow}
\usepackage{amsmath}
\usepackage{color,soul}
\usepackage[font=footnotesize, justification=raggedright,singlelinecheck=false]{caption}
\usepackage[colorlinks=true, linkcolor=blue, urlcolor=blue, citecolor=blue]{hyperref}

\begin{document}

\title{Anomalous Criticality of Absorbing Phase Transition toward Jamming}

\author{He-Da Wang}
\affiliation{National Laboratory of Solid State Microstructures,  Nanjing University,  Nanjing 210093, China}
\affiliation{Collaborative Innovation Center of Advanced Microstructures and School of Physics,  Nanjing University,  Nanjing 210093, China}
\affiliation{Jiangsu Physical Science Research Center, Nanjing 210093, China}

\author{Bo Wang}
\affiliation{National Laboratory of Solid State Microstructures,  Nanjing University,  Nanjing 210093, China}
\affiliation{Collaborative Innovation Center of Advanced Microstructures and School of Physics,  Nanjing University,  Nanjing 210093, China}
\affiliation{Jiangsu Physical Science Research Center, Nanjing 210093, China}

\author{Hao-Zheng Gu}
\affiliation{National Laboratory of Solid State Microstructures,  Nanjing University,  Nanjing 210093, China}
\affiliation{Collaborative Innovation Center of Advanced Microstructures and School of Physics,  Nanjing University,  Nanjing 210093, China}
\affiliation{Jiangsu Physical Science Research Center, Nanjing 210093, China}

\author{Qun-Li Lei}
\email{lql@nju.edu.cn}
\affiliation{National Laboratory of Solid State Microstructures,  Nanjing University,  Nanjing 210093, China}
\affiliation{Collaborative Innovation Center of Advanced Microstructures and School of Physics,  Nanjing University,  Nanjing 210093, China}
\affiliation{Jiangsu Physical Science Research Center, Nanjing 210093, China}

\author{Yu-Qiang Ma}
\email{myqiang@nju.edu.cn}
\affiliation{National Laboratory of Solid State Microstructures,  Nanjing University,  Nanjing 210093, China}
\affiliation{Collaborative Innovation Center of Advanced Microstructures and School of Physics,  Nanjing University,  Nanjing 210093, China}
\affiliation{Jiangsu Physical Science Research Center, Nanjing 210093, China}
\affiliation{Hefei National Laboratory, Shanghai 201315, China}


\date{\today}

\begin{abstract}
\textbf{Abstract:}
The jamming transition has traditionally been regarded as a geometric transition governed by static contact networks. Recently, dynamic phase transitions of athermal particles under periodic shear have offered a new perspective on this problem, leading to the conjecture that the jamming transition corresponds to an absorbing-state transition belonging to the conserved directed percolation (C-DP) or Manna universality class. Here, by re-examining the biased random organization (BRO) models on which this conjecture is based, we uncover anomalous criticality at high density. In disordered packings, we find a dynamic transition from absorbing to active-glass states with continuously varying dynamic critical exponents. Near the jamming point, quenched disorder arising from heterogeneous contact networks induces Griffiths effects, smearing the critical point into a pseudo-critical regime. In close-packed crystals, C-DP-like dynamic criticality is restored. We propose a field-theoretic description incorporating dynamical slowing down and quenched disorder to account for this anomalous criticality, providing a unified framework that links glassiness, jamming, and dynamic criticality.
\end{abstract}

\maketitle

\subparagraph{Introduction}
The jamming transition,  a transition from a soft state to a rigid state, widely exists in soft matter systems~\cite{liu2010jamming,behringer2018physics}. The static packing of spheres, especially its random close packing (RCP) state~\cite{berryman1983random}, provides an ideal model to study the jamming transition, although RCP was found to be protocol-dependent~\cite{torquato2000random,aste2005geometrical}.  

Another path to tackle the jamming problem is from the dynamic side. It was first found that when suspended colloidal particles are subjected to periodic shearing, a reversible to irreversible transition, or an absorbing phase transition, can occur upon increasing system density~\cite{corte2008random,reichhardt2023reversible}. This dynamic phase transition can be  captured by a minimal model, i.e., the random organization (RO) model~\cite{corte2008random,tjhung2015hyperuniform,
hexner2015hyperuniformity} which has been proven to belong to conserved directed percolation (C-DP) or Manna universality class~\cite{menon2009universality,torquato2018hyperuniform,2019Hydrodynamics}.  
As the shearing amplitude becomes smaller, the critical packing fraction  $\varphi_c$  increases and approach the jamming point~\cite{schreck2013particle,milz2013connecting,
reichhardt2023reversible, regev2013onset,regev2015reversibility,
kawasaki2016macroscopic,
nagasawa2019classification,
ghosh2022coupled,das2020unified,
pan2023review,deng2024jamming}. Recently, it was shown that for the biased random organization (BRO) model, a modified RO model in which the displacements of particles are not fully random but repulsive, the dynamic critical point converges to the RCP state as the displacement size (corresponding to the shearing amplitude) vanishes $(\varepsilon \rightarrow 0)$  in dimensions $d = 1$ to $5$~\cite{wilken2021random,wilken2023dynamical}. This leads to an intriguing conjecture that the jamming transition corresponds to a dynamic phase transition of the C-DP universality class~\cite{wilken2021random,wilken2023dynamical,ness2020absorbing}.  Nevertheless, near the jamming point, the  gaps (contact susceptibility) between adjacent particles are heterogeneously distributed and statically imprinted~\cite{charbonneau2015jamming,maiti2014free}. This naturally induces quenched disorder which may  significantly influence the dynamic criticality based on the \emph{Harris criterion}~\cite{harris1974effect,vojta2005critical,vojta2009infinite}. 

\begin{figure*}[!t] 
	\centering
	\resizebox{180mm}{!}{\includegraphics[trim=0.0in 0.0in 0.0in 0.0in]{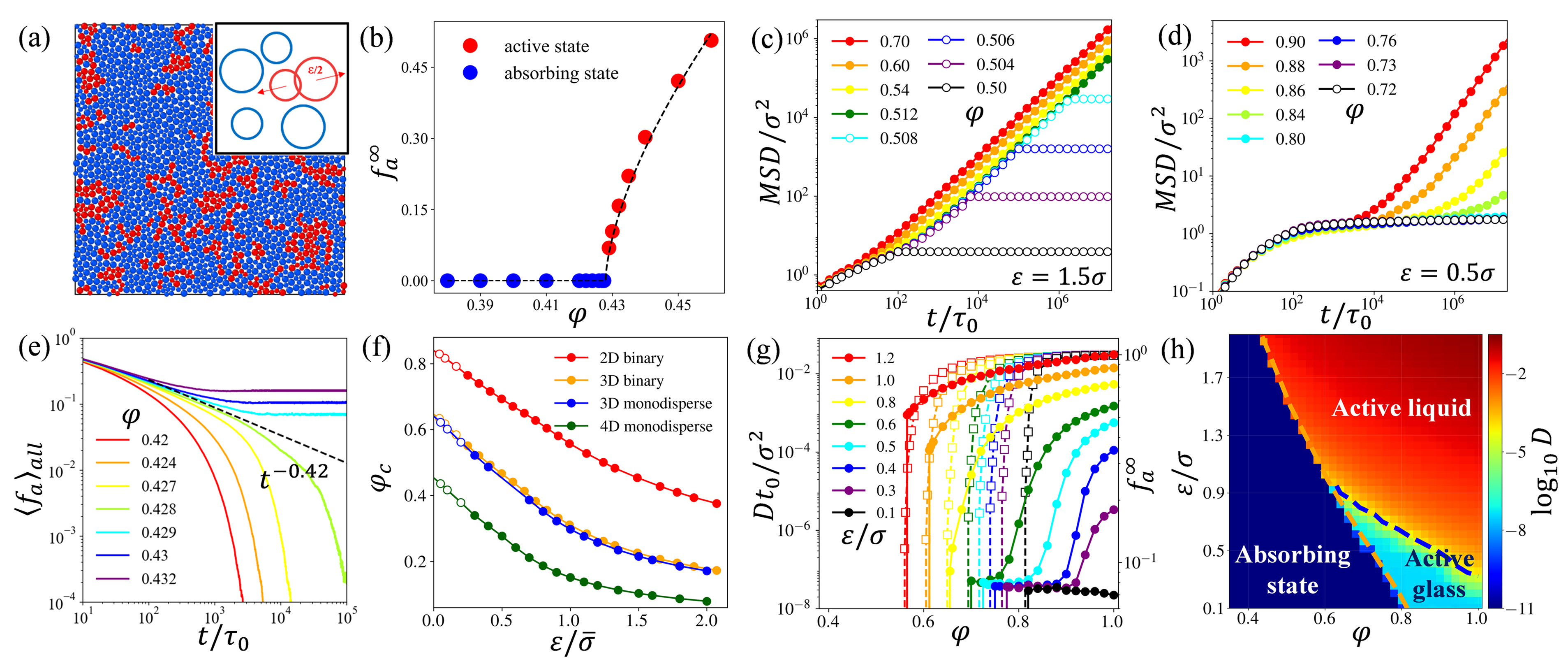} }
	\caption{  
		(a) Schematic of the conserved BRO model: overlapping particles (red) become active and take pairwise repulsive displacements. 
		(b) Order parameter $f_a^{\infty}$ as a function of packing fraction $\varphi$ at $\varepsilon=2.0\sigma$. 
		(c-d) Mean square displacement $MSD(t)$ for systems under absorbing to active liquid transition ($\varepsilon=1.5$) and absorbing to active glass transition  ($\varepsilon=0.5$), where open/solid symbols represent the absorbing/active states. 
		(e) Evolution of $\left \langle f_a \right \rangle_{all}$ for systems at $\varepsilon=2.0\sigma$ at various $\varphi$. Here, $N=65536$. 
		(f) Critical packing fractions $\varphi_c$ for systems in different dimension. Here, the solid symbols represent the normal critical points. The open symbols represent 'dirty' critical points due to the Griffiths effects. $\overline{\sigma}$ denotes the average particle diameter. 
		(g) Diffusion coefficients $D$ (solid line) and $f_a^{\infty}$ (dotted line) as functions of $\varphi$ at various $\varepsilon$. 
		(h) Dynamic phase diagram on the diffusion coefficient $D$. The yellow line marks the absorbing-active transition, while the blue line indicates the 'yielding transition'~\cite{regev2015reversibility,reichhardt2023reversible,
hima2014experimental,kawasaki2016macroscopic,nagasawa2019classification}. 
	}\label{fig1.phase_phi}
\end{figure*}

In this work, we perform a comprehensive study of the dynamic criticality of the BRO model in the high density regime and find  anomalous critical behavior different from C-DP universality class. In dense disordered systems, we identify an absorbing to active-glass transition which is accompanied by  varying  dynamic critical exponents. With a further increase of $\varphi_c$ by decreasing $\epsilon$,  Griffiths effects emerge as a result of quenched disorder in particle contact network, smoothing the dynamic transition and broadens the critical point into a pseudo-critical regime. Close to the jamming point ($\varepsilon \rightarrow 0$),  the dynamic transition becomes equivalent to directed percolation (DP) with quenched disorder.  These results indicate that while the BRO model can generate RCP configurations, the criticality of the transition is fundamentally reshaped by the RCP state. For close-packed crystal structures,  Griffiths effects are absent. The C-DP-like dynamic criticality is restored. We  propose a modified C-DP field theory, incorporating the effects of dynamic slowdown and quenched disorder to explain the anomalies, which provides an unified framework that  links glassiness, jamming, and dynamic criticality. Our findings shed light on the dynamic criticality of more complicated disordered systems, like the learning process of artificial neural networks~\cite{zhang2024absorbing,spigler2018jamming,anand2026emergent}.

\subparagraph{Model and simulations}
In the BRO model, $N$ particles are first randomly distributed in a hypercubic box of length $L$ and dimension $d$ at packing fraction $\varphi$.  At each discrete time step, isolated (non-overlapping) particles are inactive and stay put; overlapping particles become active and receive opposite repulsive displacements of magnitude  $\varepsilon/2$ (Fig.~\ref{fig1.phase_phi}(a))~\cite{hexner2017noise,wilken2021random,wilken2023dynamical,galliano2023two}. We consider two BRO dynamics: with or without local center-of-mass conservation (CMC)~\cite{hexner2017noise} (see End Matter).  Since both models exhibit similar results, we present CMC-conserved results  in the main text, while leaving non-conserved results in Supplemental Materials~\cite{supmat}.  Both  monodisperse  and symmetric bidisperse particle systems are considered. In the latter, the particle diameter ratio is $\sigma_1/\sigma_2 = 1:\sqrt2$ and the small diameter is set as the unit length $\sigma$.

\begin{figure*}[!t] 
	\centering
	\resizebox{180mm}{!}{ \includegraphics[width=\textwidth]{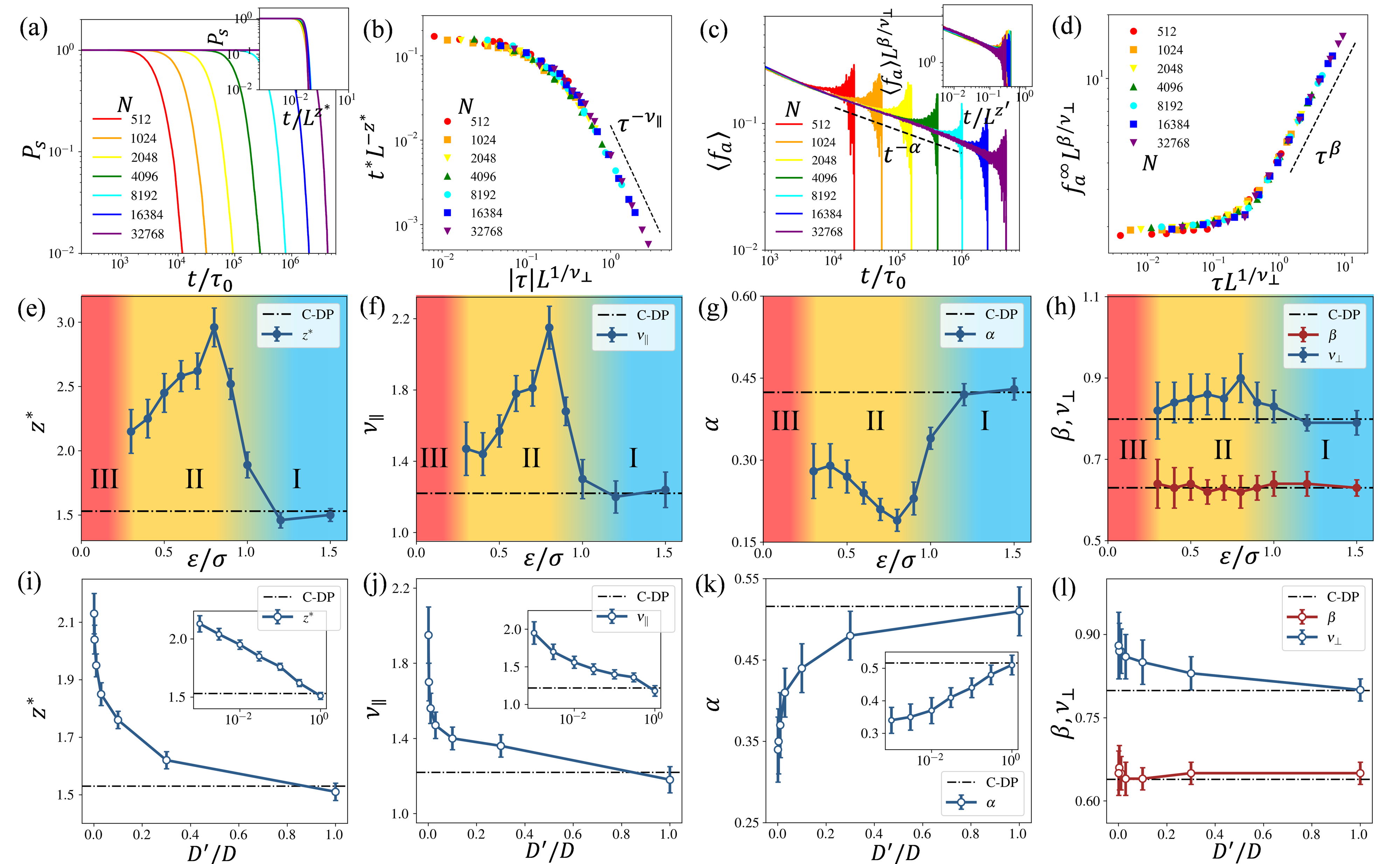}}
	\caption{
		({a}-{d}) Finite size scaling of the survival probability $P_s(t)$, characteristic life time $t^*$, survival activity $\left \langle f_a \right \rangle(t)$ and $f_a^{\infty}$ at $\varepsilon=0.8\sigma$, $\varphi_c=0.6499$.
		({e}-{h}) Critical exponents of $z^*$, $\nu_{\parallel}$, $\alpha$, $\beta$ and $\nu_{\perp}$ as functions of $\varepsilon$, from which one can distinguish regime I (light blue), regime II (light orange), regime III (light red). The dotted lines represent the values in the C-DP universality class~\cite{lubeck2004universal,henkel2008non}.
		({i}-{l}) Same as ({e}--{h}) but for the field theory model.  Note that the reference value of  $\alpha$ is $0.51$ for field simulation~\cite{dornic2005integration,ramasco2004numerical} instead of $0.42$ in Manna models. It may due to coarse-graining of the fields, or a reflection of true asymptotic scaling~\cite{dornic2005integration,ramasco2004numerical,lubeck2003universal}.
	}\label{fig2.exponent}
\end{figure*}

To examine the dynamic criticality, we define the distance from the critical point as   $\tau=(\varphi-\varphi_c)/\varphi_c$, and measure the relaxation of survival activity $\left \langle f_a \right \rangle(\tau, t)$, the overall activity $\left \langle f_a \right \rangle_{all}(\tau, t)$ and  the survival probability  $P_s(\tau, t)$, in  an ensemble of dynamic trials starting from random configurations. At the critical point ($\tau=0$), $\left \langle f_a \right \rangle(t)$ and $\left \langle f_a \right \rangle_{all}(t)$ would decay as  $t^{-\alpha}$ (see  Fig.~\ref{fig1.phase_phi}{(e)} and Fig.~\ref{fig2.exponent}(c)). The equilibrium value $f_a^{\infty} \equiv  \left \langle f_a \right \rangle (t \rightarrow \infty)$ serves as the order parameter of the absorbing transition Fig.~\ref{fig1.phase_phi}(b). 
Through the finite-size scaling of $\left \langle f_a \right \rangle$, $\left \langle f_a \right \rangle_{all}$ and $P_s$,  we obtain a set of critical exponents  $\beta$, $\nu_{\parallel}$, $\alpha$, $z^*$ and $\nu_{\perp}$  (Fig.~\ref{fig2.exponent} and End Matter).

\subparagraph{General behaviors}

We first study the 2D bidisperse case in which $\varepsilon$ is larger than particle size $\sigma$. Under this condition, the motion of active particles is not constrained by neighboring particles and  the absorbing transition is essentially a non-local searching process for a non-overlapping configuration. (Fig.~\ref{fig1.phase_phi}{(c)}).  The obtained critical exponents as functions of $\varepsilon$ are shown in Regime I of Fig.~\ref{fig2.exponent}(e-h), which are consistent with the C-DP universality class.   

As $\varepsilon$ decreases below $\sigma$, $\varphi_c$  increases to high density. Due to the repulsive displacement in BRO model,  particles are caged by their neighbors in the active state (Fig.~\ref{fig1.phase_phi}(d)), a characteristic of the glass-forming liquids~\cite{debenedetti2001supercooled,2025Structural}. In these states, the diffusion coefficient $D$ of active particles  can be as low as $10^{-6} \sigma^2/\tau_0$ while the activity $f_a^{\infty}$ remains large (Fig.~\ref{fig1.phase_phi}(g)). In Fig.~\ref{fig1.phase_phi}{(h)}, we plot the  phase diagram of $D$, where the active-glass state (more precisely, active glass-forming liquid state) is highlighted in the cyan regime.  Interestingly, as $\varphi$ further increases at fixed $\varepsilon$, the system becomes diffusive again, which corresponds to the yielding transition of granular systems under oscillatory shearing~\cite{regev2015reversibility,reichhardt2023reversible,
hima2014experimental,kawasaki2016macroscopic,nagasawa2019classification}. 
For the absorbing to active-glass transition, we find substantial variations in $\nu_{\perp}$, $\alpha$, and $z$ (Regime II in Fig.~\ref{fig2.exponent}(e-h)). This suggests that glassy behavior significantly influences the dynamic criticality of the system. 
 

Further decreasing $\varepsilon$ below $0.3\sigma$ (Regime III), the critical density (if well defined)  approaches that of RCP state, in which cage-breaking events become unlikely and the topology of particle network remains unchanged in the active state.  In this pre-jamming regime, we observe the failure of standard finite size scaling and the uncertainty of the critical point (Fig.~S4-S7~\cite{supmat}).

The above anomalous critical phenomena are also observed in 3D and 4D  monodisperse or bidisperse systems (Fig.~S2, Fig.~S8-S11~\cite{supmat}). In Fig.~\ref{fig1.phase_phi}(f)), we plot $\varphi_c$ as a function of $\varepsilon$, where the uncertain $\varphi_c$ in Regime III are represented by open symbols. By extrapolating these data to $\varepsilon \rightarrow 0$, we obtain $\varphi_c=0.838$ and 0.645 in 2D and 3D bidisperse systems, $\varphi_c=$0.640 and 0.453 in 3D and 4D monodisperse systems, which are consistent with the  RCP states $\varphi_{\rm RCP}$~\cite{berryman1983random,skoge2006packing,2023Estimating}, and in line with~\cite{wilken2021random,wilken2023dynamical}. Note that in 3D monodisperse systems, crystallization prevents accurate determination of the critical point for $0.4\sigma<\varepsilon<0.7\sigma$ (see Fig.~\ref{fig1.phase_phi}{(f)} and Fig.~S3~\cite{supmat}).

\begin{figure}[!t] 
\centering
\resizebox{90mm}{!}{\includegraphics[width=\columnwidth]{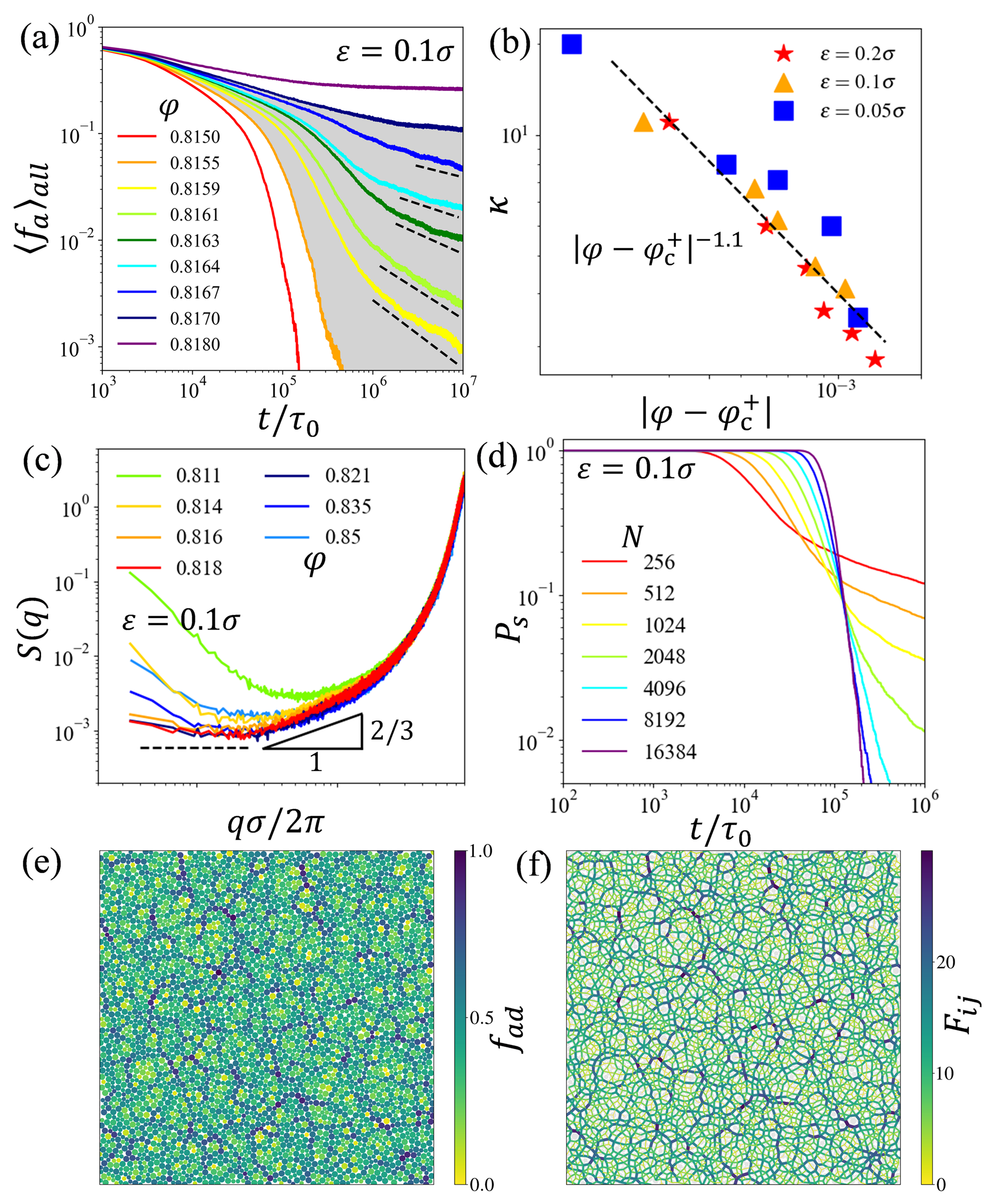} }
\caption{	  
		({a}) Evolution of $\left \langle f_a \right \rangle_{all}$ for different $\varphi$, with the Griffiths regime shaded in gray ($N=8192$).
		({b}) Measured $\kappa$ index (Eq.~(\ref{kappa})) as a function of $\varphi$ for different $\varepsilon$. The dirty critical densities are $\varphi^+_c = 0.7916$, $0.8170$, and $0.8297$ for $\varepsilon=0.2\sigma$, $0.1\sigma$, and $0.05\sigma$, respectively.
		({d}) Finite-size scaling of the survival probability $P_s(t)$.
		({e}) Spatial distributions of the normalized activity duration $f_{ad}$.
		({c}) Structure factor for 2D BRO systems $S(q)$. 
		({f}) Spatial distribution of the force chains for the same configuration in ({d}).
        For ({e}, {f})  $\varepsilon = 0.1\sigma$ and $\varphi = 0.8152$.
}  \label{fig3.griffith}
\end{figure}

\subparagraph{Field Theory of Absorbing to active-glass transition}
It is known that the C-DP universality class is described by a non-conserved field $ \rho({\bf{r}},t)$ (active particle density field) coupled with a conserved field $ \psi({\bf{r}},t)$ (total density field)~\cite{rossi2000universality,pastor2000field}, namely,
\begin{eqnarray}
\frac{\partial \rho(t)}{\partial t} &=& D \nabla^2 \rho +a \rho-b \rho^2 + \omega \rho \psi + \sigma_n \sqrt{ \rho} ~ \xi(t)   \label{Adiff_eq}  \\
\frac{\partial \psi(t)}{\partial t} &=& D' \nabla^2 \rho,   \label{diff_eq}
\end{eqnarray}
typically with $D=D'$, where  Eq.~(\ref{Adiff_eq}) describes the activation and de-activation processes, with $\omega \rho \psi$ accounting for the coupling  between two fields, and $\sigma_n \sqrt{ \rho} \xi(t) $ the multiplicative noise term. Eq.~(\ref{diff_eq}) reflects that the evolution of $ \psi({\bf{r}},t)$ is driven by the spreading of active particles. 

In the active-glass state,  $\rho(t)$ field may still spread spatially through local contact-mediated activation, but the translational diffusion of active particles is strongly suppressed. A rigorous field treatment of glass state involves a retarded, nonlocal memory kernel~\cite{archer2006dynamical,gotze1998essentials}. Nevertheless,  particles eventually become diffusive after typical relaxation time. In this sense, the active glass state might still be described by Eq.~(\ref{diff_eq}) with suppressed diffusion constant $D' \ll D$ in the large time scale. 
We therefore directly simulate  Eq.~(\ref{Adiff_eq}-\ref{diff_eq})  with  $D'/D$ ranging  from 1 to $10^{-3}$~\cite{Dornic2005} (see Supplemental Material~\cite{supmat}). We find that decreasing  $D'/D$ can  roughly reproduce the variation in dynamic critical exponents observed in BRO model (Fig.~\ref{fig2.exponent}(i-l) and Fig.~S12-S13~\cite{supmat})). Nevertheless, the observed monotonic changes in the critical exponents differ from the BRO model, indicating other effects related {to} jamming may play roles as $\epsilon \rightarrow 0$.

In Ref.~\cite{le2015exact}, it was argued that the $D'/D$ ratio does not affect the C-DP universality class. It remains an open question whether the anomalous critical exponents observed here are crossover phenomena due to subtle finite-size effects~\cite{Kaywiese} or other factors not considered in previous theory~\cite{le2015exact}. 


\subparagraph{Griffiths effects in pre-jamming state}
In the pre-jamming Regime III, different from the conventional absorbing state transition (Fig.\ref{fig1.phase_phi}(e)),  we find $\left \langle f_a \right \rangle_{all}$  obeys a continuous power law 
\begin{eqnarray}
\left \langle f_a \right \rangle_{all}(t) \sim t^{-d/\kappa}        \label{kappa}   
\end{eqnarray}
with the $\varphi$-dependent index $\kappa \sim |\varphi-\varphi^+_c|^{-h}$ in a density range $\varphi_c^-<\varphi<\varphi^+_c$\cite{vojta2005critical}, as shown  in Fig.~\ref{fig3.griffith}(a-b). Here  $h$ is an exponent on the `dirty' critical point $\varphi^+_c$~\cite{vojta2005critical}.  The existence of this pseudo-critical regime rather than a critical point, is a characteristic of the Griffiths effects which is usually caused by  quenched disorders~\cite{munoz2010griffiths, moreira1996critical,webman1998dynamical,
hooyberghs2004absorbing,vojta2005critical,vojta2006nonequilibrium}.   In Fig.~\ref{fig3.griffith}(b), we plot the measured $\kappa$ as a function of  $|\varphi-\varphi^+_c|$ for  systems under different $\varepsilon$. The fitting roughly gives $h=1.1$~\cite{supmat}.  

In the pre-jamming phase, the finite-size effect on $P_s(t)$ near criticality is also reversed, i.e., at the same density near the transition, small systems stay longer in the active state than large systems (Fig.~\ref{fig3.griffith}{(d)}). This anomaly is nevertheless consistent with the fact that small systems are more likely to be jammed ~\cite{goodrich2012finite,xu2005random},  suggesting that close-packing significantly influences the criticality of absorbing phase transition. 

The Griffiths effect is a reflection of the violation of Harris criterion, i.e., uncorrelated weak quenched disorder is irrelevant for phase transition when  $d\nu_{\perp}> 2$, since for C-DP universality class, $d\nu_{\perp}< 2$ for $d<4$. In the BRO model near the jamming point, the quenched disorder originates  from heterogeneously distributed particles gaps (or free volume)~\cite{charbonneau2015jamming,maiti2014free} which hardly changes (Fig.~S14~\cite{supmat}). Particles with smaller particle gaps are more susceptible be activated. This contact susceptibility are imprinted in the pre-jamming state. This quenched disorder is reflected in the spatial distribution of normalized activity duration $f_{ad}$ (see Fig.~\ref{fig3.griffith}(e) and End Matter), which resembles the force chains of the jammed state~\cite{radjai1996force} if mapping the hard-sphere BRO configurations onto an soft-sphere model (see Fig.~\ref{fig3.griffith}(f)) and Supplemental Materials~\cite{supmat}).  The spatial  correlation coefficients between the  activity duration and force chains $\mathbf{F}_{ij}$ are as high as $C \simeq 0.882$ and $0.925$ for systems with $\varepsilon = 0.1\sigma$ and $0.01\sigma$, respectively. Thus particles that  are more easily activated are precisely those that carry larger forces in the same configurations of soft spheres.


The origin of  quenched disorder  can  be understood theoretically. In the jamming limit ($\varepsilon \rightarrow 0$),  Position of particles  can be decomposed into an inherent structure  (the local minimum of potential energy landscape) and vibrational displacement  around it~\cite{stillinger1982hidden}.
Correspondingly, the density field  can be decomposed into an intrinsic density field contribution $ \psi_{\rm IS}({\bf r}, t)$ and a vibrational contribution $\delta \psi_{\rm vb}({\bf r}, t)$, i.e.,$\psi({\bf r}, t) =  \psi_{\rm IS}({\bf r}, t)  + \delta \psi_{\rm vb}({\bf r}, t) $. In the jamming limit, the inhomogeneous $ \psi_{\rm IS}({\bf r}, t)$ freezes ($D' \rightarrow 0$). Its influence is much stronger and persistent than its vibrational counterpart. Thus $\psi({\bf r}, t)   \simeq   \psi_{\rm IS}({\bf r})$. Moreover, $\omega$ in Eq.~(\ref{Adiff_eq})  represents the susceptibility of the active density field to  density fields $\omega$ , which is determined by the contact susceptibility in BRO model. In the jamming limit, $\omega$ becomes singular and spatial heterogeneous. Thus, Eqs.~(\ref{Adiff_eq}-\ref{diff_eq}) reduce to
\begin{eqnarray}
	\frac{\partial \rho(t)}{\partial t} &=& D \nabla^2 \rho +a \rho + \omega({\bf r}) \psi_{\rm IS}({\bf r}) \rho-b \rho^2  + \sigma_n \sqrt{ \rho}~\xi(t) ~~~~~~~~
	\end{eqnarray}  
which is the field theory of directed percolation (DP) with quenched disorder $\omega({\bf r}) \psi_{\rm IS}({\bf r})$~\cite{hooyberghs2004absorbing,moreira1996critical,
vojta2005critical}. 

The classical contact model on a heterogeneous network is the simplest realization of DP~\cite{1974Contact,jensen1992critical}. To further confirm the above argument, we simulate the contact model on the same jammed networks with infection rate proportional to proportional to $f_{ad}$ (see Supplemental Material~\cite{supmat}).  It exhibits dynamic heterogeneity and anomalous finite-size scaling similar to the high-density  BRO (Fig.~S15~\cite{supmat}), proving that the critical behaviors of near-jammed  BRO model indeed reduces to  DP with quenched disorders. 

It is well-known that RO/BRO models at low critical densities are  critical hyperuniform states characterized by structure factor  $S(q) \sim q^{\theta}$ with $\theta=0.45$ (2D) and $0.25$ (3D)~\cite{tjhung2015hyperuniform,hexner2015hyperuniformity}.  It was argued that $\theta$ remains unchanged when the system approaches the jammed point $\varepsilon \rightarrow 0$~\cite{wilken2021random,wilken2023dynamical}. 
For 2D BRO model, we confirm $\theta=0.45$ when $\varepsilon >0.5 \sigma$ (Fig.~S16~\cite{supmat}), consistent with~\cite{wilken2021random,wilken2023dynamical}. However,  when approaching the jamming point ($\varepsilon =0.1 \sigma$), notable deviation can be found (Fig.~\ref{fig3.griffith}(d)), which is  in line with~\cite{dam2026bidispersity,galliano2026glass}, and further prove that the jamming transition is uncorrelated with C-DP universality class. For 3D  systems,  the  exponent $\theta=0.25$ is too small for us to be accurately determined in the critical jammed state (Fig.~S17 ~\cite{supmat}).

\subparagraph{Discussion and Conclusion}

In summary, we investigate the critical behavior of BRO model in high density monodisperse and bidisperse systems from $d=2$ to 4, and find  anomalous critical phenomena, like varying dynamic critical exponents and Griffiths effects, in contrast to standard C-DP universality class. Our analysis indicates that although the BRO model can generate RCP configurations, the criticality of the dynamic transition is fundamentally reshaped by the geometric rigidity (force chain) of the RCP state.  As a comparison, for the close-packed crystal phase, where quenched disorder and Griffiths effects are absent, C-DP-like dynamic criticality is restored (see End Matter).  All these critical phenomena can be explained by a unified field theory. We believe that our work could enrich further studies of non-equilibrium glass systems~\cite{nishikawa2021relaxation,galliano2026glass}, as well as the dynamic criticality of  artificial neural networks~\cite{spigler2018jamming,
anand2026emergent,zhang2024absorbing}.

\begin{acknowledgments} 
\subparagraph{Acknowledgments:} The authors are grateful to Ran Ni, S. Wilken, K. J. Wiese, L. Berthier and many others for helpful discussions. This work is supported by  the National Natural Science Foundation of China (No.~12347102, 12275127),  the Natural Science Foundation of Jiangsu Province (No. BK20233001, No. BK20250058),  the Fundamental and Interdisciplinary Disciplines Breakthrough Plan of the Ministry of Education of China (No. JYB2025XDXM502), Program for Innovative Talents and Entrepreneur in Jiangsu, the Fundamental Research Funds for the Central Universities (0204-14380249, KG202501), the Innovation Program for Quantum Science and Technology (No. 2024ZD0300101). The simulations are performed on the High-Performance Computing Center of Collaborative Innovation Center of Advanced Microstructures, the High-Performance Computing Center (HPCC) of Nanjing University. 
\end{acknowledgments} 

\bibliography{ref}

\section{End Matter}

\subparagraph{CMC conserved and non-conserved BRO model}
In the CMC conserved  BRO model, the displacements of a pair of overlapping particles have the same magnitude randomly chosen from a uniform range of $[0, \varepsilon/2]$.  If a particle overlaps with multiple particles, these displacements (vectors) are superimposed~\cite{hexner2017noise,galliano2023two}. To realize  CMC non-conserved dynamics, we adopt two methods. The first one is based on ~\cite{wilken2021random, wilken2023dynamical}, where the displacement magnitude of each overlapped particle is randomly chosen from $[0, \varepsilon/2]$, and the direction of displacement of the particle is from the mass center of overlapped clusters to the center of this particle. The second one is based on the introduction of 20\%   random displacement for each particle in the original the CMD conserved dynamics as done in~\cite{wilken2021random}. See~\cite{supmat} for details.

\subparagraph{Finite-size scaling of absorbing phase transition}

The survival activity of the system at time $t$ is defined as
\begin{equation}
	\left \langle f_a  \right \rangle(\tau, t)=\frac{<N_a(\tau, t)>_a}{N}
\end{equation}
where $N_a(t)$ is the number of active particles at time step $t$. Here, $\tau=(\varphi-\varphi_c)/\varphi_c$ is the distance to the critical point.  Note that  $<>_a$ here indicates that the average is only on trials that remain active up to time $t$. We can further define the overall activity of the system averaged over all trials
\begin{equation}
	\left \langle f_a \right \rangle_{all}(\tau, t)=\frac{<N_a(t)>_{all}}{N} = \left \langle f_a  \right \rangle(\tau, t) P_s(\tau, t)
\end{equation}
where   $P_s(\tau, t)$ is the  survival probability of the active phase up to time $t$, from which the characteristic life time $t^*$ of the trials can be extracted.  For typical second-order absorbing phase transitions, $\left \langle f_a \right \rangle(t)$ should satisfy the finite-size scaling relationship~\cite{henkel2008non,lubeck2003universal,lei2021barrier}
\begin{eqnarray}
\left \langle f_a  \right \rangle(t,\tau,L)\propto L^{-\beta/\nu_{\perp}}   \mathcal{F}(t L^{-z' }, \tau L^{1/\nu_{\perp}}) 
\end{eqnarray}
where $\mathcal{F}(x,y)$ is the scaling function satisfying $ \mathcal{F}(x,0)\sim x^{-\alpha}$ for $x \ll 1$ and $ \mathcal{F}(x,0)\sim const.$ for $x \gg 1$.  In a fully active state, the decay of $\left \langle f_a  \right \rangle(t,0,L)$ is independent of $L$ at an earlier stage, which leads to $z'=\beta/(\nu_{\perp}{\alpha})$. On the other hand, $f_a^{\infty}$ satisfies
\begin{equation}
f_a^{\infty}(\tau,L) \propto L^{-\beta/\nu_{\perp}}  \mathcal{F}(\infty,\tau L^{1/\nu_{\perp}})
\end{equation}
where $ \mathcal{F}(\infty, y) \sim y^\beta$ for $y\gg 1$ and $ \mathcal{F}(\infty, y) \sim const.$ for $y \ll 1$. Furthermore, the survival probability $P_s(\tau, t)$ of the active phase up to time $t$ should satisfy
\begin{eqnarray}
P_s(t, \tau, L)\propto  \mathcal{P}(t L^{-z^*}, \tau L^{1/\nu_{\perp}})  \label{FS_Ps}
\end{eqnarray}
where $ \mathcal{P}(x,0) \sim 1$ for $x \ll 1$ and $ \mathcal{P}(x,0) \rightarrow 0$ for $x\gg 1$.  From Eq.~(\ref{FS_Ps}), we can also obtain the finite size scaling relationship for the characteristic life time of activity,
\begin{eqnarray}
t^*  \propto  L^{z^*} \mathcal{T}( \tau L^{1/\nu_{\perp}}) 
\end{eqnarray}
which is the time at which the survival probability decays to 1\%,  i.e., $\mathcal{P}(t^*, \tau, L)  =0.01$. Here, $ \mathcal{T}(x) \sim const.$ for $x \ll 1$ and $ \mathcal{T}(x) \rightarrow x^{-\nu_{\parallel}}$ for $x\gg 1$. By definition, $z=\nu_{\parallel}/\nu_{\perp}$ is the dynamic critical exponent. For a typical absorbing phase transition, one should expect $z=z'=z^*$.

\subparagraph{Activity duration distribution}
To visualize the spatial heterogeneity of particle activation susceptibility, we first initiate the system from a random configuration and let it evolve under the BRO dynamics  up to a time limit of $10^6$ steps. Subsequently, we apply $100$ perturbation--relaxation cycles. In each cycle, we apply perturbation to each particle to restart the dynamics until the system either falls back into an absorbing state up to $10^5$ steps. Then, we record the total time $t_a(i)$ during which particle $i$ remains active in that subsequent evolution. The normalized activity duration $f_{ad}(i)$ for particle $i$ is defined as
\begin{equation}
	f_{ad}(i)= \left \langle \frac{t_a(i)}{t_{\max}} \right \rangle,
\end{equation}	
where $t_{\max}$ is the active life time of the system. For each configuration, the $f_{ad}(i)$ distribution is obtained from the accumulated active times over 100 cycles.

\begin{table}
	\renewcommand{\arraystretch}{1.5}		
	\setlength{\tabcolsep}{1.5mm}
	\begin{tabular}{l|llllll}
		\hline
		$\varepsilon$ & $\beta$ & $\nu_\perp$ & $\alpha$ & $\nu_\parallel$ & $z^{*}$ & $z$\\ \hline
		0.01$\sigma$ & {0.64(2)} & {0.79(3)} & {0.52(4)} & {1.08(6)} & {1.49(6)} &  {1.37(9)} \\ 
		0.2 $\sigma$  & {0.63(3)} & {0.79(2)} &  {0.53(3)} &  {1.07(5)} & {1.46(7)} &  {1.35(7)} \\
		C-DP & 0.639 & 0.799 & {0.52*} & 1.225 & 1.533 & 1.533 \\ \hline
	\end{tabular}
	\vspace{10pt}
	\caption{
		Critical exponents measured in 2D crystal systems under conserved BRO dynamics. Reference values for the C-DP universality class are from~\cite{lubeck2004universal,henkel2008non}. {*Note that considering the homogeneous nature of crystal phase,  the $\alpha$ reference value was obtained based on natural homogeneous initial configurations instead of random configurations~\cite{lei2021barrier,basu2012fixed,lee2013comment}.}  }\label{tab1} 
\end{table}	

\subparagraph{Dynamic criticality in crystal phase}
For a defect-free crystal, after coarse-graining over the lattice  scale, the inherent density field is homogeneous, i.e., $\psi_{\rm IS}({\bf r})=\psi_{\rm IS}^0$. The density activation coupling $\omega$ also becomes homogeneous singular, i.e., $\omega(\bf r)= \omega_0$, which can be proved by the Gaussian like gap distribution in crystal phase (Fig.~S14~\cite{supmat}).  Hence the quenched disorder and corresponding Griffiths effect are absent.  The fluctuation of $\psi(t)$ is solely determined by $\delta \psi_{\rm vb}(t)$ which is connected to local velocity field ${\mathbf v}(t)$, i.e., ${\delta} \psi_{\rm vb}(t)/{\partial t} ={-} \psi_0  \nabla {\cdot} {\mathbf v}(t)$~\cite{chaikin1995principles}. Generally, for overdamped systems, $v_i= - \mu_{ik}\partial_j \sigma_{kj}$, where $\mu_{ik}$ is mobility tensor and ${\mathbf  \sigma}_{kj}$ is the effective stress tensor. For non-equilibrium crystal phases near the critical point, ${\mathbf  \sigma}_{kj}$ can be generally written as a function of $\rho$ in the lowest order, i.e., $\sigma_{kj}  \simeq \rho  Q_{kj} $, where  $Q_{kj}$ is a constant tensor reflecting the  anisotropy of crystal phase. This gives the C-DP-like field equation 
\begin{eqnarray}
	&&\frac{\partial \rho(t)}{\partial t} = D \nabla^2 \rho +a' \rho-b \rho^2 + \omega_0 \rho  \delta  \psi_{\rm vb} (t)  + \sigma_n \sqrt{ \rho} ~ \xi(t) ~~~~~~  \label{Cry_eq}  \\
	&&\frac{\partial  \delta  \psi_{\rm vb}(t) }{\partial t} = D'_{ij} \partial_i \partial_j   {\rho} (t),   \label{Cry_eq2}  
\end{eqnarray}
where $a'=a+\omega_0 \psi_{\rm IS}^0$ and  $D'_{ij}= \psi_0  \mu_{ik} Q_{kj}$  is the diffusion coefficient tensor. FOr hexagonal crystal, diffusion tensor  is isotropic, i.e., $D'_{ij}=D'_0 \delta_{ij}$. Then, Eq.~(\ref{Cry_eq}-\ref{Cry_eq2}) returns to the field equation for the C-DP universality class~\cite{rossi2000universality,pastor2000field}.  In Table.~\ref*{tab1}  and {Fig.~S18}~\cite{supmat}, we give the measured critical exponents for crystal phase. We find the critical exponents $\beta$, $\nu_{\perp}$, $\alpha$ are consistent with C-DP universality class, in line with~\cite{galliano2023two}, but  $\nu_{\parallel}$ and  $z$ exhibit notable deviations. We  speculate that this may due to other effects like high-order anisotropic effect of crystal structures or phononic-like collective motion~\cite{guo2026diffusion} that are beyond the description of the current field theory.

\end{document}